\documentclass[preprint,12pt]{aastex}

\shorttitle{Magnetic Fields in a Massive Star-Forming Filament}
\shortauthors{Qiu et al.}
\begin{document}

\title{From Poloidal to Toroidal: Detection of Well-ordered Magnetic Field in High-mass Proto-cluster G35.2$-$0.74\,N}

\author{Keping Qiu\altaffilmark{1,2}, Qizhou Zhang\altaffilmark{3}, Karl M. Menten\altaffilmark{4}, Hauyu B. Liu\altaffilmark{5}, Ya-Wen Tang\altaffilmark{5}}
\email{kpqiu@nju.edu.cn}
\altaffiltext{1}{School of Astronomy and Space Science, Nanjing University, 22 Hankou Road, Nanjing 210093, China}
\altaffiltext{2}{Key Laboratory of Modern Astronomy and Astrophysics (Nanjing University), Ministry of Education, Nanjing 210093, China}
\altaffiltext{3}{Harvard-Smithsonian Center for Astrophysics, 60 Garden Street, Cambridge, MA 02138, U.S.A.}
\altaffiltext{4}{Max-Planck-Institut f\"{u}r Radioastronomie, Auf dem H\"{u}gel 69, 53121 Bonn, Germany}
\altaffiltext{5}{Academia Sinica Institute of Astronomy and Astrophysics, P. O. Box 23-141, Taipei, 106 Taiwan}

\begin{abstract}
We report on detection of an ordered magnetic field (B field) threading a massive star-forming clump in the molecular cloud G35.2$-$0.74, using Submillimeter Array observations of polarized dust emission. Thanks to the sensitive and high-angular-resolution observations, we are able to resolve the morphology of the B field in the plane of sky and detect a great turn of $90^{\circ}$ in the B field direction: Over the northern part of the clump, where a velocity gradient is evident, the B field is largely aligned with the long axis of the clump, whereas in the southern part, where the velocity field appears relatively uniform, the B field is slightly pinched with its mean direction perpendicular to the clump elongation. We suggest that the clump forms as its parent cloud collapses more along the large scale B field. In this process, the northern part carries over most of the angular momentum, forming a fast rotating system, and pulls the B field into a toroidal configuration. In contrast, the southern part is not significantly rotating and the B field remains in a poloidal configuration. A statistical analysis of the observed polarization dispersion yields a B field strength of $\sim1$~mG, a turbulent-to-magnetic energy ratio of order unity, and a mass-to-magnetic flux ratio of $\sim2$--3 in units of the critical value. Detailed calculations support our hypothesis that the B field in the northern part is being rotationally distorted. Our observations, in conjunction with early single-dish data, suggest that the B field may play a critical role in the formation of the dense clump, whereas rotation and turbulence could also be important in further dynamical evolution of the clump. The observations also provide evidence for a wide-angle outflow driven from a strongly rotating region whose B field is largely toroidal.
\end{abstract}


\section{Introduction} \label{intro}
The role of magnetic fields in the evolution of molecular clouds and the formation of stars has long been subject to great debate \citep[see][for a review]{Crutcher12}. In a classic model of low-mass star formation, molecular clouds are supported by magnetic fields, and stay in subcritical states, evolving quasi-statically; ambipolar diffusion induces the formation of supercritical cores which dynamically collapse to form stars \citep{Shu87,Basu94,Mouschovias06}. The other view is that molecular clouds are short lived, dynamically evolving, and producing stars rapidly \citep{Elmegreen00,Hartmann12}, with magnetic field being implicitly weak, or not really appreciable in cloud evolution and star formation; Mac Low and Klessen (2004) further strongly argue that supersonic turbulence, instead of magnetic field, supports molecular clouds and regulates star formation. On the other hand, our understanding of high-mass star formation is far less clear, but the competing views on the role of magnetic field/turbulence are equally, if not more strongly debated. Recent magnetohydrodynamic (MHD) simulations suggest that magnetic fields are dynamically important in high-mass star formation, in particular at suppressing complete fragmentation and creating bipolar outflows  \citep{Banerjee07,Peters11,Hennebelle11,Commercon11,Seifried12,Myers13}.

The ``magnetic support'' model predicts a well-ordered or, in the extreme case, uniform magnetic field permeating a molecular cloud \citep{Ostriker01}. Since there is an increased support against gravity in the direction perpendicular to the magnetic field compared to the direction parallel to it, the cloud contracts more along the field, forming flattened cores orthogonal to the mean direction of the field \citep{Matsumoto04,Tassis09}. In contrast, in the ``weak field'' or ``turbulent support'' model, magnetic field is expected to show an irregular and even chaotic morphology due to overwhelming turbulent twisting \citep{Ostriker01,Padoan01}. Therefore, mapping the morphology of magnetic field provides a straightforward method to distinguish between the two competing paradigms, or to provide insights into the possibility of a scenario where both magnetic field and turbulence are important. Focusing on massive clumps or cores with a typical size scale of 0.1~pc and a distance of a few kpc, high-angular-resolution observations of polarized dust emission are needed to spatially resolve the magnetic field morphology. Submillimeter Array (SMA) observations have been playing a major role in recent studies of this kind, though the observations are still limited to a small number of case studies \citep[e.g.,][]{Girart09,Tang13,Liu13,Girart13}. Here we present an SMA\footnote[6]{The SMA is a joint project between the Smithsonian Astrophysical Observatory and the Academia Sinica Institute of Astronomy and Astrophysics and is funded by the Smithsonian Institution and the Academia Sinica.} study of a massive cluster-forming clump which shows an elongated morphology projected in the plane of sky, and thus defines an axis ready to be compared to the direction of the magnetic field.

The targeted clump (hereafter G35.2N) lies in the northern part of G35.2$-$0.74, a molecular cloud first discovered by Brown et al. (1982) and located at a distance of 2.19~kpc \citep{Zhang09}. It is associated with the IRAS source 18566+0136, which has a total luminosity of $3\times10^4$~$L_{\odot}$ \citep{Dent89,Sanchez-Monge13}. Early molecular line observations revealed a velocity gradient along the long axis of the clump as well as a large scale bipolar outflow approximately orthogonal to the clump elongation \citep{Dent85,Little85,Brebner87}. G35.2N was thus interpreted as a rotating interstellar disk or toroid. More detailed studies of the molecular outflow were presented by Gibbs et al. (2003) and Birks et al. (2006). Most recently, Zhang et al. (2013) presented SOFIA-FORCAST mid-infrared observations of G35.2N and modeled the object as a single high-mass protostar forming by an ordered and symmetric collapse of a massive core with a radius of 0.1~pc. However, both Atacama Large Millimeter/submillimeter Array (ALMA) cycle 0 observations \citep{Sanchez-Monge13} and the SMA data presented here reveal an apparently filamentary and highly fragmented structure for the dense gas on a 0.15~pc scale. We further infer a well-ordered B field from sensitive observations of the polarized dust emission and find that the B field morphology correlates with the dense gas kinematics.

\section{Observations and Data Reduction} \label{obs}
The observations were carried out with the SMA from 2010 to 2012, using three array configuration under excellent weather conditions. Detailed information on the observations, including the observing dates, array configurations, number of available antennas, atmospheric opacities at 225~GHz, and various calibrators, is presented in Table~\ref{table1}. For the Subcompact and Extended observations, the 345~GHz receiver was tuned to cover roughly 332--336~GHz in the lower sideband and 344--348~GHz in the upper sideband. For the Compact observations, the frequency setup covers about 333.5--337.5~GHz in the lower sideband and 345.5--349.5~GHz in the upper sideband. For all the observations, the correlator was configured to have a uniform spectral resolution of 812.5~kHz ($\sim$0.7~km\,s$^{-1}$).

We performed basic data calibration, including bandpass, time dependent gain, and flux calibration, with the IDL MIR package, and output the data into MIRIAD for further processing. The intrinsic instrumental polarization (i.e., leakage) was removed to a 0.1\% accuracy \citep{Marrone08} with the MIRIAD task GPCAL. For each sideband, a pseudo-continuum data set was created from spectral line-free channels using the MIRIAD task UVLIN. The calibrated visibilities were jointly imaged to make Stokes $I$, $Q$, and $U$ maps. We performed self-calibration with the continuum data in Stokes $I$, and applied the solutions to both continuum and spectral line data in Stokes $I$, $Q$, $U$. Finally, the Stokes $I$, $Q$, and $U$ maps were combined to produce maps of the polarized emission intensity, the fractional polarization, and the polarization angle, using the MIRIAD task IMPOL. No significant polarization is detected in molecular line emission, so only Stokes $I$ maps are presented for spectral lines.

\section{Results} \label{result}
\subsection{Dust Emission and Magnetic Field}
\begin{figure}
\epsscale{.4} \plotone{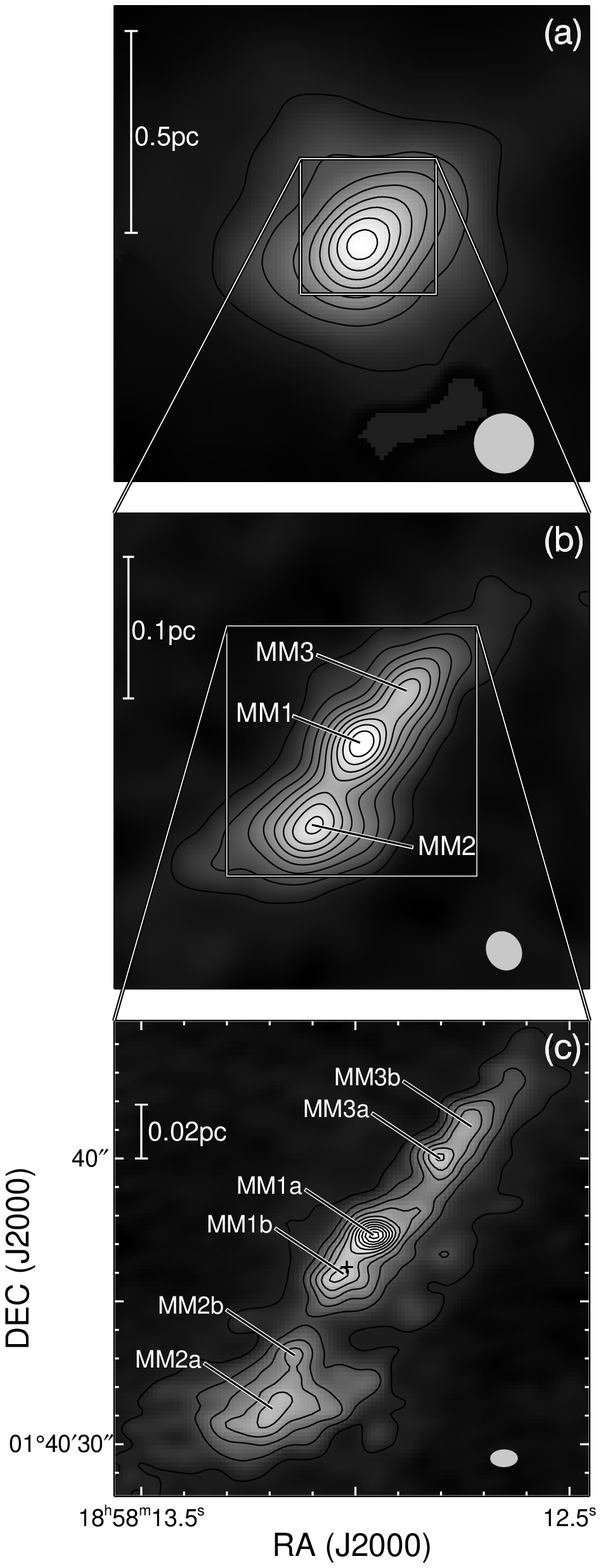}
\caption{The total dust continuum emission, shown in both inverse gray scale and contours, at 850~$\mu$m observed with the JCMT SCUBA and at 880~$\mu$m with the SMA. (a) The SCUBA map, with the starting and spacing contour levels of 1~Jy\,beam$^{-1}$. (b) The SMA map made from the Subcompact and Compact data, with the contour levels of $0.04\times(1, 2, 3, ...)^{1.5}$~Jy\,beam$^{-1}$. (c) The SMA map made from all the data available, with the contour levels of $0.015\times(1, 2, 3, ...)^{1.4}$~Jy\,beam$^{-1}$. A plus sign marks the position of a weak radio continuum source \citep{Gibb03}. Negative emissions are invisible since their absolute levels are all below the lowest contour levels shown here. Hereafter, a filled ellipse in the lower right or left of a panel depicts the beam size at FWHM. \label{cont}}
\end{figure}

Figure \ref{cont} shows the total dust emission on a variety of size scales. In the James Cleak Maxwell Telescope\footnote[7]{The James Clerk Maxwell Telescope is operated by the Joint Astronomy Centre on behalf of the Science and Technology Facilities Council of the United Kingdom, the Netherlands Organisation for Scientific Research, and the National Research Council of Canada.} (JCMT) SCUBA 850~$\mu$m map with a $15''$ resolution (Figure \ref{cont}a), the emission unveils a $\sim$0.5~pc, slightly elongated clump, which shows hierarchical fragmentation in our SMA observations. In the SMA 880~$\mu$m map made by combining the Subcompact and Compact observations (Figure \ref{cont}b), the inner clump splits into three cores, MM1--3, each with a diameter of order 0.05~pc. Convolving the SMA map to the SCUBA beam and comparing its peak flux ($\sim$6~Jy) to that of the SCUBA map ($\sim$8.5~Jy), the SMA observations recover approximately 70\% of the total flux. Figure \ref{cont}c shows the SMA map made from all the observations spread over five tracks (Table \ref{table1}). With a uniform weighting of the data, we obtain an angular resolution of  $1.\!''0\times0.\!''6$, and resolve each of the three cores into at least two condensations with diameters of 0.01--0.02~pc. These condensations are distributed approximately along the major axis of the clump, but in MM2, there are minor emission peaks to the west. MM1b approximately coincides with an unresolved radio source first detected at 8.5~GHz \citep{Gibb03}. S\'{a}nchez-Monge et al. (2013) presented the ALMA 860~$\mu$m continuum map at a $0.\!''4$ resolution. Our SMA map shown in Figure \ref{cont}c is in general consistent with the ALMA observations.

\begin{figure}
\epsscale{.55} \plotone{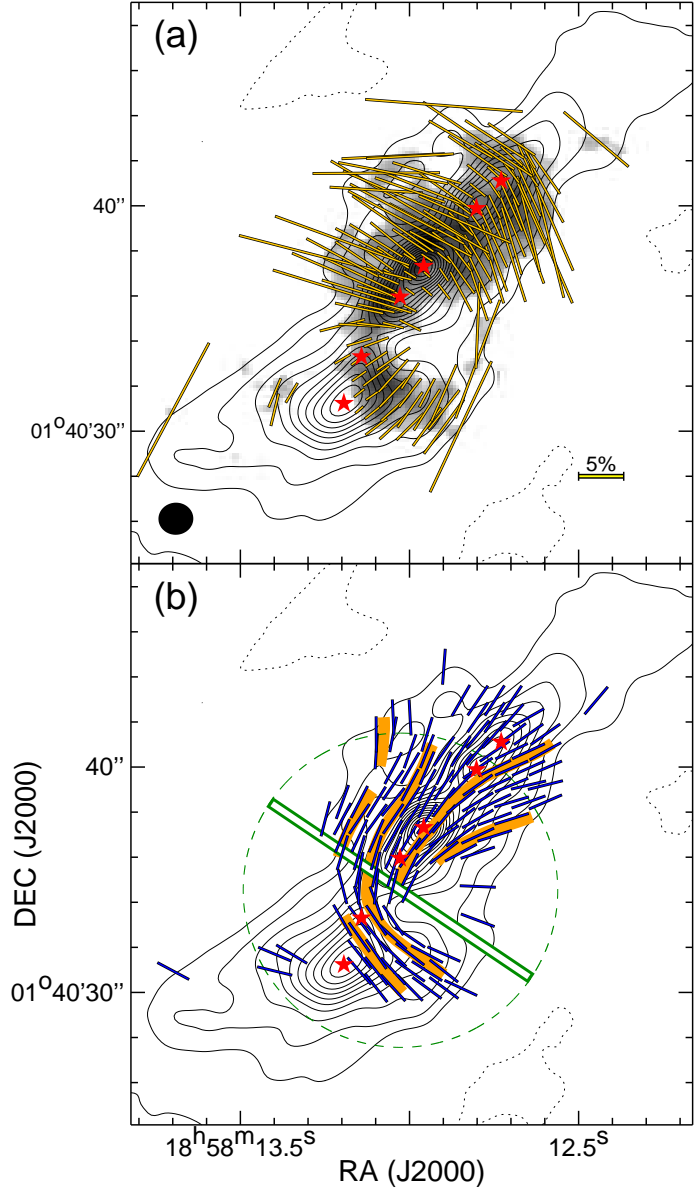}
\caption{(a): The total dust emission (Stokes $I$) shown in contours (solid for positive and dotted for negative emission) and the linearly polarized emission ($\sqrt{Q^2+U^2-\sigma^2}$) detected with signal-to-noise ratios $\geq3$ are shown in gray scale; the contour levels are $\pm0.03\times(1, 2, 3, ...)^{1.5}$~Jy\,beam$^{-1}$. Yellow segments indicate the directions of the linear polarization, with their lengths proportional to the fractional degree of the polarization; a scale bar in the lower right corresponds to a polarization degree of 5\%. Star symbols mark the peak positions of the six dust condensations. (b): Contours and star symbols are the same as in (a). Blue segments with an arbitrary length show the B field directions, deduced by rotating the polarization directions by $90^{\circ}$. A large green bar indicates the direction of the B field obtained from early JCMT observations of the polarized dust emission at $760~\mu$m \citep{Vallee00}; a dashed circle marks the $14''$ beam of the observations. Brown curves are drawn following a method proposed by Li et al. (2010), and outline the continuous variation in the direction of the B field at representative positions. \label{pol_mag}}
\end{figure}

To achieve the highest possible sensitivity, which is desirable for a polarization study, we jointly imaged all the SMA observations under a natural weighting, which results in an r.m.s. noise level, $\sigma$, of $1.5$~mJy\,beam$^{-1}$. Strong polarization of the dust emission (signal-to-noise ratios greater than 3) is detected toward MM1, MM3, and part of MM2 (Figure \ref{pol_mag}a), and thus the inferred B field in the plane of sky is well resolved (Figure \ref{pol_mag}b). We detect a clearly ordered B field threading the clump. Over the northern two cores MM1 and MM3, the B field direction is overwhelmingly aligned with the long axis of the clump, whereas approaching the southern core MM2, the B field direction drastically changes by about $90^{\circ}$ so that it is perpendicular to the clump elongation. Interestingly, early single-dish polarimetric observations with a $14''$ beam revealed a mean direction at a position angle of $56^{\circ}$ for the B field permeating the parent cloud of the clump \citep[][also see Figure \ref{pol_mag}b]{Vallee00}. Therefore, over MM1 and MM3, the B field in the dense clump is approximately perpendicular to the mean direction of the large scale B field, while toward MM2, the B field in the clump roughly follows that in the cloud. In addition, to the southwest of MM1 and to the east of MM2, there are  B field segments inclined to the direction of the cloud B field. Even for the northern part of the clump, the direction of the B field on the edges of the clump tends to deviate more from the clump major axis.

\subsection{Molecular Line Emission}
Our SMA observations cover a large number of spectral lines tracing molecular gas under a variety of physical conditions. No significant polarization is detected in these lines, but they allow a detailed investigation of gas kinematics.

\subsubsection{Velocity gradient in high-density tracing molecular lines} \label{kinematics}
\begin{figure}
\epsscale{.85} \plotone{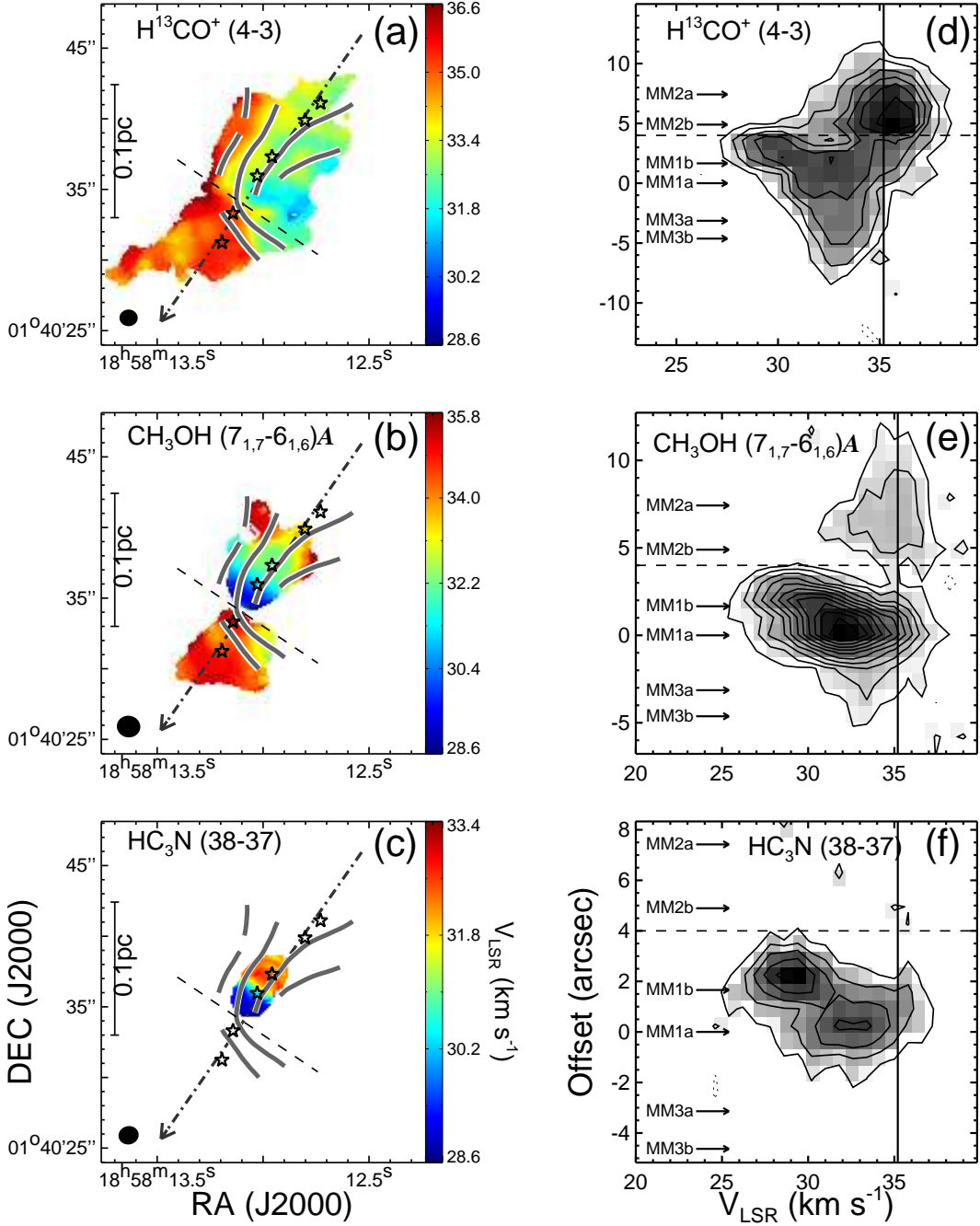}
\caption{(a--c) Color images show the first moment (intensity weighted velocity) maps of the H$^{13}$CO (4--3), CH$_3$OH $(7_{1,7}$--$6_{1,6})\,A$, and HC$_3$N (38--37) lines, respectively. A dash-dotted arrow delineates the $PV$ cut at a position angle of $-35^{\circ}$ and points to the direction of increasing offsets. A dashed line divides the clump into two parts which are discussed in the text (Sections \ref{kinematics} and \ref{winding}). Other symbols (stars and curves) are the same as in Figure \ref{pol_mag}b. (d--f) $PV$ diagrams in the three spectral lines shown in both gray scale and contours (solid for positive and dotted for negative emission). The contour levels are $\pm0.15\times(1, 2, 3, ...)^{1.5}$~Jy\,beam$^{-1}$. A dashed line marks the position of the dashed line in panels (a--c). A vertical line indicates the systemic velocity of G35.2-0.74 \citep{Roman09}. The positions of the six condensations are denoted in each panel. \label{gradient}}
\end{figure}

We investigate the kinematics of the dense gas by examining the first moment (intensity weighted velocity) maps (Figures \ref{gradient}a--c) and position-velocity ($PV$) diagrams (Figures \ref{gradient}d--f) of various high-density tracing molecular lines.

In Figure \ref{gradient}, the H$^{13}$CO$^+$ (4--3) emission traces the bulk dense gas of the clump seen in dust continuum. The most remarkable feature in the first moment map of the emission is a velocity shear lying between MM1 and MM2. From the $PV$ diagram cut along the major axis of the clump (position angle $-35^{\circ}$), MM2, which has a mean velocity close to the systemic velocity of the ambient cloud G35.2$-$0.74 \citep{Roman09}, can be readily distinguished from the rest of the clump, MM1 and MM3, and a velocity gradient is clearly seen between the latter two cores' locations. The CH$_3$OH $(7_{1,7}$--$6_{1,6})\,A$ emission preferentially traces the inner regions of the clump, and reveals an overall velocity field consistent with that seen in H$^{13}$CO$^+$ (4--3). The HC$_3$N (38--37) line has an upper level energy of 324~K above the ground, and probes the two condensations, MM1a and MM1b, embedded within MM1. There is a clear velocity gradient across the two condensations in the first moment map and $PV$ diagram of the HC$_3$N (38--37) emission. We also investigated our SMA 230~GHz observations, and found a similar velocity structure along the clump.

The dense gas kinematics and the B field morphology both suggest that the clump can be understood as a two-component structure: the northern part, which consists of MM1 and MM3, exhibits a velocity gradient of order 50~km\,s$^{-1}$~pc$^{-1}$, and the mean direction of the B field is aligned with the long axis of the clump; the southern part, i.e., MM2, does not show a clear velocity gradient, and the B field direction is perpendicular to the clump axis, such that it is aligned with the B field in the cloud.

\subsubsection{Hot molecular cores and a wide-angle outflow} \label{HMC-outflow}
\begin{figure}
\epsscale{1} \plotone{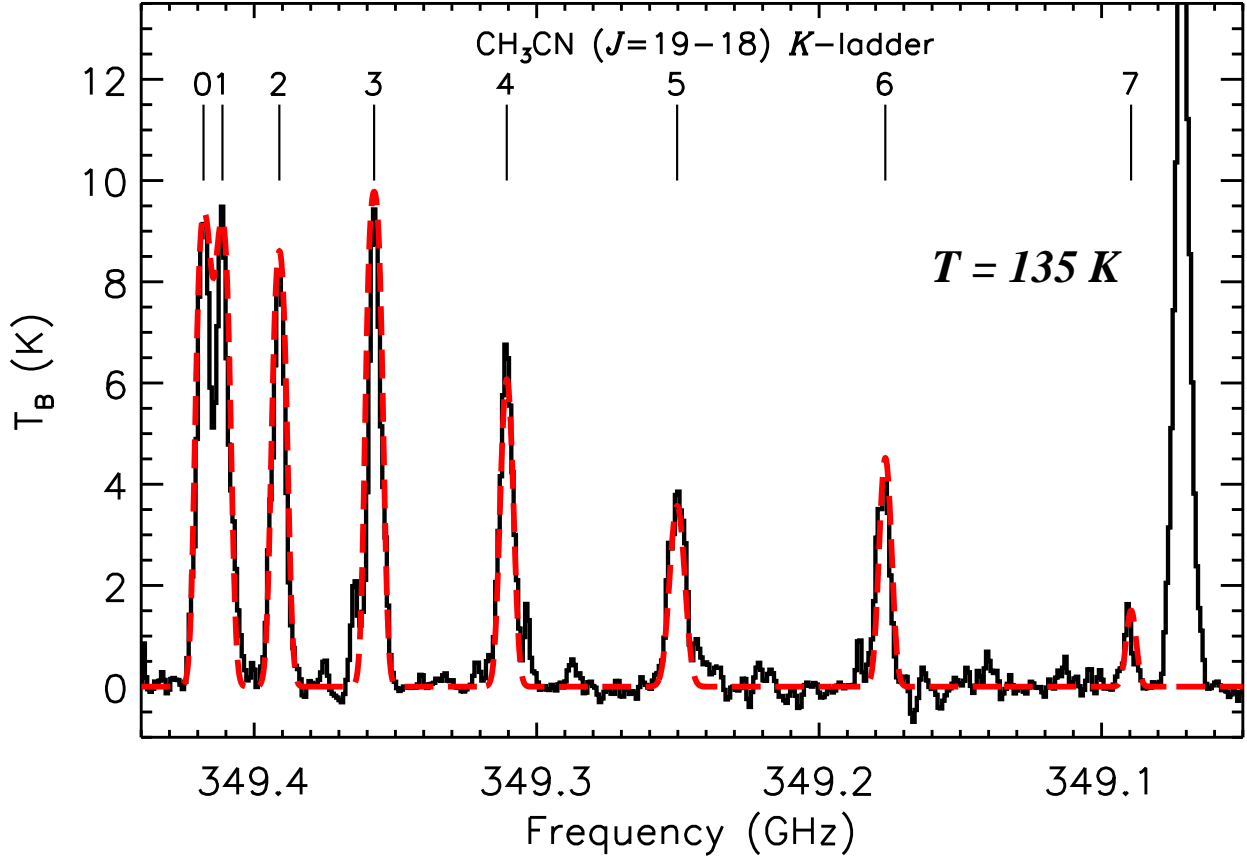}
\caption{Solid histograms show the observed spectra of the CH$_3$CN (19--18) $K$-ladder, overlaid with the best-fit LTE model, which yields a gas temperature of 135 K, shown in a dashed curve. A very strong spectral line with a (sky) frequency of $\sim349.07$~GHz is CH$_3$OH ($14_{1,13}$--$14_{0,14}$)$A$, and is irrelevant to the fitting to the CH$_3$CN lines. \label{ch3cn}}
\end{figure}

The two dust condensations embedded within MM1 (i.e., MM1a,b) are the brightest in the clump, and show the richest molecular lines. MM1b coincides with (within $0.\!''5$) a faint and spatially unresolved radio source detected at 8.5~GHz \citep{Gibb03}. Following Qiu \& Zhang (2009) and Qiu et al. (2011b), we perform a local thermodynamical equilibrium (LTE) fitting to the $K$-ladder of the CH$_3$CN (19--18) emission that is detected toward these two condensations. In Figure \ref{ch3cn}, the best-fit model agrees well with the observation, and yields a temperature of $135^{+35}_{-15}$~K, indicating that the two sources are hot molecular cores (HMCs) and confirming that the clump is actively forming high-mass stars.

\begin{figure}
\epsscale{1} \plotone{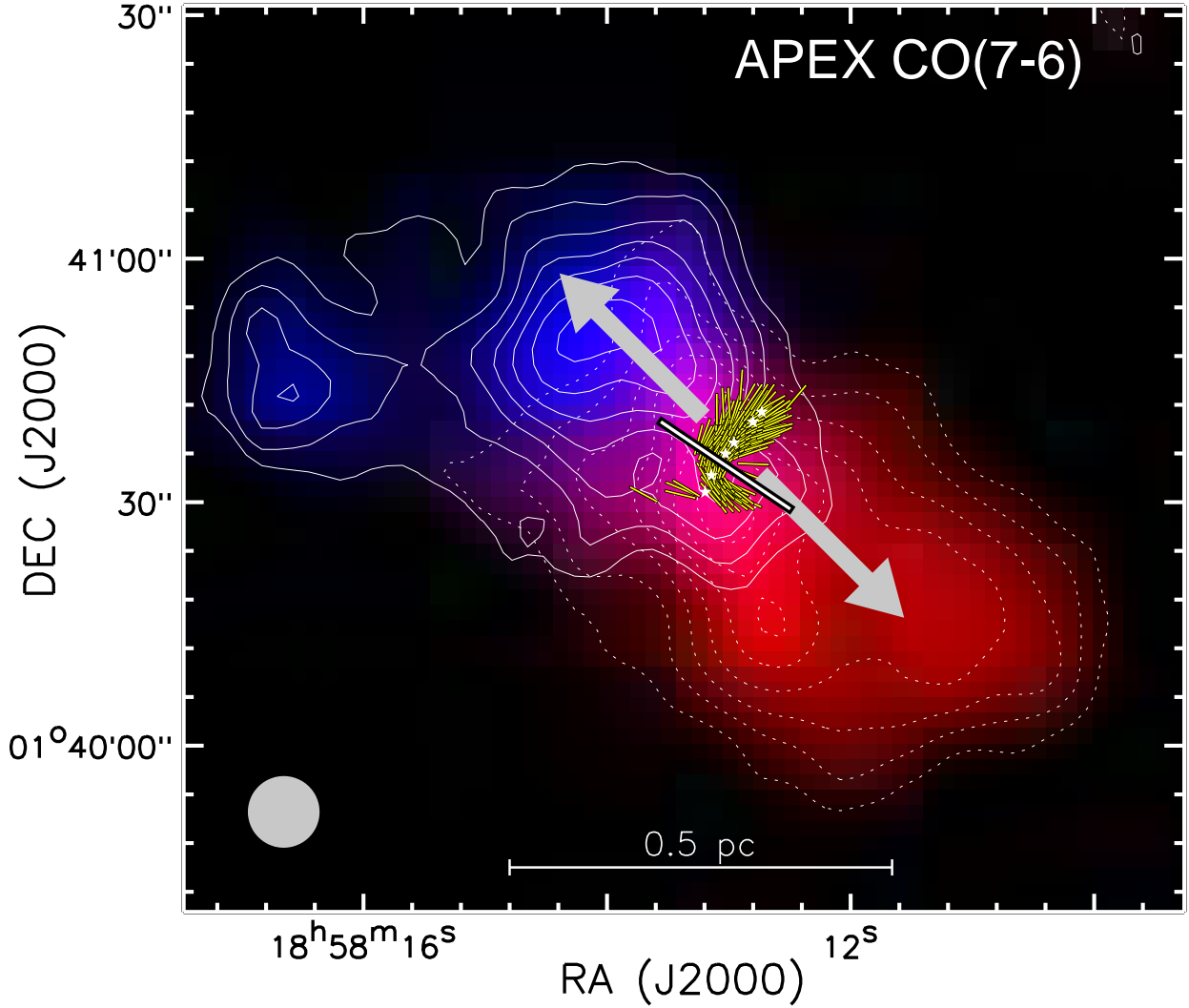}
\caption{The velocity integrated CO (7--6) emission observed with the APEX telescope. The blue background and solid contours show the blueshifted emission, and the red background and dotted contours show the redshifted emission. Two arrows approximately mark the outflow axis. Other symbols are the same as in Figure \ref{pol_mag}b. \label{outflow}}
\end{figure}

The CO (3--2) line is covered in our SMA observations. However, due to the nearly equatorial position of the source and a limited number of antennas available in the interferometer, the CO maps are heavily affected by sidelobe and missing flux issues, and do not allow a reliable investigation of the outflow structure. We instead show our Atacama Pathfinder Experiment (APEX) CO (7--6) observations, which reveal a wide-angle bipolar outflow centered on the two HMCs (Figure \ref{outflow}). The outflow morphology is in general agreement with previous low-$J$ CO observations \citep{Gibb03,Birks06}. A more detailed study of the outflow nature in G35.2N, using multi-frequency observations, will be presented in a forthcoming paper. Here we stress that the outflow axis is approximately perpendicular to the clump elongation (thus also perpendicular to the B field in the northern part of the clump).

\section{Discussion}
\subsection{The B field strength and its significance compared to turbulence and gravity}
\begin{figure}
\epsscale{.8} \plotone{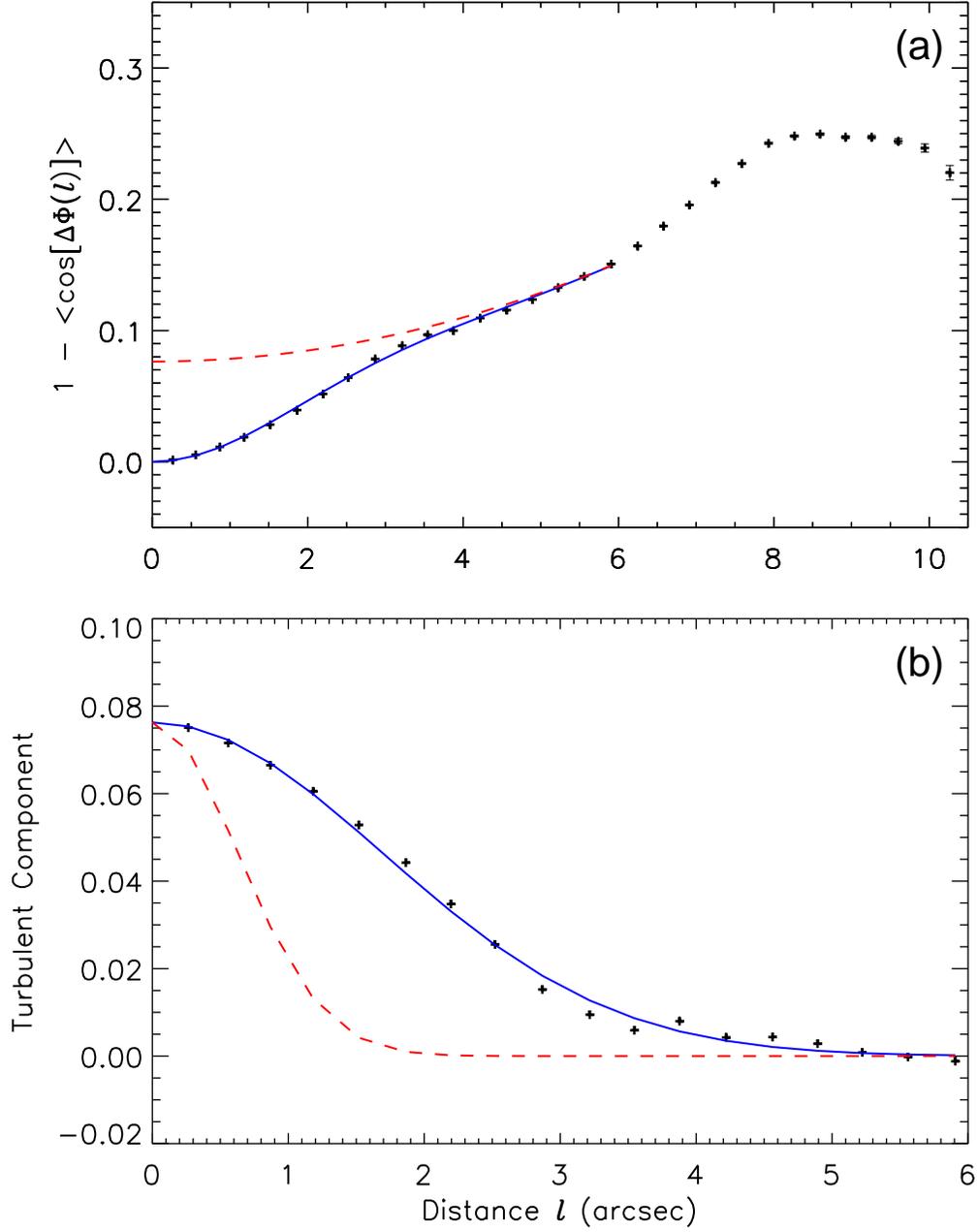}
\caption{(a) Plus signs show the measured dispersion function, $1-\langle{\rm cos}[\Delta{\Phi}(l)]\rangle$. For most data points, the error bars are too small to recognize (barely visible at $l>9''$). A solid curve shows the fitted function, $\frac{1}{N}\frac{{\langle}B_t^2\rangle}{{\langle}B_0^2\rangle}\times[1-e^{-l^2/2({\delta}^2+2W^2)}]+a_2'l^2$, for $l\lesssim6''$; a dashed curve visualizes the sum of the integrated turbulent contribution and the large scale contribution, i.e., $\frac{1}{N}\frac{{\langle}B_t^2\rangle}{{\langle}B_0^2\rangle}+a_2'l^2$. (b) The correlated component of the dispersion function: data points (shown in plus signs) are derived by subtracting the measured dispersion function from $\frac{1}{N}\frac{{\langle}B_t^2\rangle}{{\langle}B_0^2\rangle}+a_2'l^2$ ; the fitted $\frac{1}{N}\frac{{\langle}B_t^2\rangle}{{\langle}B_0^2\rangle}e^{-l^2/2({\delta}^2+2W^2)}$ is shown in a solid curve. A dashed curve visualizes the correlation solely due to the beam of the observation. \label{dispersion}}
\end{figure}

We detect a structured B field threading the clump (Figure \ref{pol_mag}). The observed P.A. dispersion is attributed to both ordered and random perturbations to the B field, and the latter is probably dominated by turbulence. But dynamical processes related to cluster formation could perturb the B field in a manner rendering some disordered to random variations. We perform a statistical analysis of the observed P.A. dispersion to quantify the random perturbations \citep{Hildebrand09,Houde09,Koch10}. Assuming the observed B field to be composed of a large-scale, ordered component, $B_0$, and a turbulent component, $B_t$, Houde et al. (2009) show that the disperse function $1-\langle{\rm cos}[\Delta{\Phi}(l)]\rangle$, which is measurable with a polarization map, can be approximately expressed as $$1-\langle{\rm cos}[\Delta{\Phi}(l)]\rangle\simeq\frac{1}{N}\frac{{\langle}B_t^2\rangle}{{\langle}B_0^2\rangle}\times[1-e^{-l^2/2({\delta}^2+2W^2)}]+a_2'l^2,$$ where $\Delta{\Phi}(l)$ is the difference in P.A. measured at two positions separated by a distance $l$ and $\langle\cdot\cdot\cdot\rangle$ denotes an average, $\delta$ is the turbulent correlation length, $W$ is the beam radius (i.e., the FWHM beam divided by $\sqrt{8{\rm ln}2}$), and $a_2'$ is the slope of the second-order term. $N$ is the number of independent turbulent cells probed by observations, and is defined by $$N=\frac{({\delta}^2+2W^2){\Delta}'}{\sqrt{2\pi}{\delta}^3},$$ where $\Delta'$ is the effective depth of the cloud along the line of sight. Here turbulence is implicitly adopted to be the only source of the random perturbations. Figure \ref{dispersion}(a) shows the measured function taking into account polarization detections with signal-to-noise ratios $>4$ in the northern part of the G35.2N clump\footnote[8]{We excludes the southern part to constrain the P.A. in the range [$0^{\circ}$, 90$^{\circ}$]}. In our observations, $W=0.\!''63$, and $\Delta'\sim8$--$22''$, which are approximately the dimensions of the clump projected on the plane of sky (Figure \ref{pol_mag}). We then fit the measured function with the above relation to solve for ${\langle}B_t^2\rangle/{\langle}B_0^2\rangle$, $\delta$, and $a_2'$, of which the former two are of our interest. Since the relation is valid when $l$ is less than a few times $W$, we fit the data points at $l\lesssim6''$. The solid curve in Figure \ref{dispersion}a shows the fitting results, and the dashed curve visualizes the sum of the integrated turbulent contribution, $[{\langle}B_t^2\rangle/{\langle}B_0^2\rangle]/{N}$, and the large-scale contribution, $a_2'l^2$. The correlated turbulent component of the dispersion function, $$\frac{1}{N}\frac{{\langle}B_t^2\rangle}{{\langle}B_0^2\rangle}e^{-l^2/2({\delta}^2+2W^2)},$$ is shown in Figure \ref{dispersion}b. With the fitting we obtain $\delta=15.4$~mpc ($1.\!''45$), which is approximately equal to the value derived in OMC-1 \citep{Houde09}), and ${\langle}B_t^2\rangle/{\langle}B_0^2\rangle=0.23$--0.64 for $N=3.0$--8.4, which depends on $\Delta'$. According to Chandrasekhar \& Fermi (1953), $$\left[\frac{{\langle}B_t^2\rangle}{{\langle}B_0^2\rangle}\right]^{1/2}\sim\frac{{\delta}V_{\rm los}}{V_{\rm A}},$$ where ${\delta}V_{\rm los}$ is the one-dimensional velocity dispersion and $V_{\rm A}=B_0/\sqrt{4\pi\rho}$ is the Alfv\'{e}n velocity at mass density $\rho$. Consequently we can estimate the energy ratio of the turbulence to the ordered B field, ${\beta}_{\rm turb}\sim3{\langle}B_t^2\rangle/{\langle}B_0^2\rangle$, to be 0.7--1.9. The ratio is somewhat overestimated considering that we are probing the B field in the plane of sky and $B_t$ is not necessarily all ascribed to turbulence. Nevertheless, the statistical analysis suggests that the ordered B field and turbulence are energetically comparable.

The strength of the ordered B field can be derived following $$B_0=\sqrt{4\pi\rho}\cdot{\delta}V_{\rm los}\cdot\left[\frac{{\langle}B_t^2\rangle}{{\langle}B_0^2\rangle}\right]^{-1/2}.$$ The H$^{13}$CO$^+$ (4--3) emission traces a structure similar to that is seen in the dust emission, and we estimate ${\delta}V_{\rm los}\sim1$~km\,s$^{-1}$ from the line-of-sight velocity dispersion of the H$^{13}$CO$^+$ line. To estimate $\rho$, we first compute the mass of the clump from its dust emission, which requires the information on the dust temperature and opacity. In our SMA observations, the clump consists of three cores, MM1--3, each of which harbors a couple of dust condensations. Toward MM2 and MM3, early NH$_3$ (1,1) and (2,2) observations deduced a rotational temperature of 20--25~K, corresponding to a kinetic temperature of 30--40~K \citep{Little85}. Toward MM1, our LTE model of the CH$_3$CN (19--18) $K$-ladder gives a temperature of 135~K for the embedded HMCs, while recent high-angular-resolution NH$_3$ observations suggest a temperature $>50$~K \citep{Codella10}. Thus we adopt a gas temperature of 50--135~K for MM1. Assuming thermal equilibrium between gas and dust, the derived gas temperature approximates the dust temperature. The dust opacity index, $\beta$, is derived by comparing the dust emission fluxes at 880~$\mu$m and 1.3~mm (the latter is obtained from our new SMA observations). The dust opacity at 880~$\mu$m is then extrapolated from 10$[\nu/(1.2 \,\rm{THz})]^{\beta}$, where $\nu$ is frequency, following Hildebrand (1983). Assuming a canonical gas-to-dust mass ratio of 100, the masses of the three cores are then computed from their dust emission fluxes (Table \ref{table2}). The total mass of the clump amounts to $\sim150$~$M_{\odot}$, and results in an averaged column density of $\sim3.7\times10^{23}$~cm$^{-2}$ over a measured area, $s$, of 200 square arc\,sec (0.015~pc$^2$) where the emission is detected with signal-to-noise ratios above 5. If we adopt a volume of $4/3(s^3/\pi)^{1/2}$, the averaged volume density is $\sim1.0\times10^6$~cm$^{-3}$. We then obtain $B_0\sim1.4$--0.9~mG, again depending on $\Delta'$. Consequently, the mass-to-magnetic flux ratio, $M/{\phi}_B$, is 1.9--3.2 in units of a critical value, $1/(2\pi\sqrt{G})$, where $G$ is the gravitational constant \citep{Nakano78}, indicating that the clump is unstable against gravitational collapse and fragmentation

\begin{figure}
\epsscale{1} \plotone{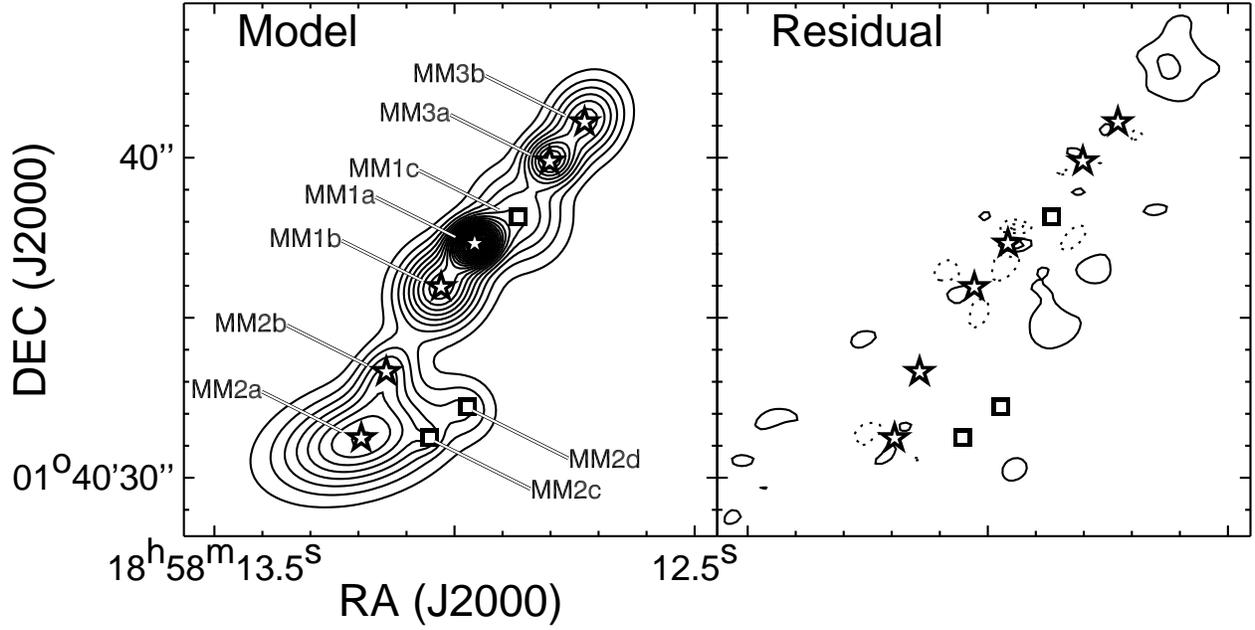}
\caption{The {\it Left} panel shows a model of the total dust emission derived from a multi-component Gaussian fitting to Figure \ref{cont}c; the starting and spacing contour levels are both 0.018~Jy\,beam$^{-1}$. Star symbols are the same as those in Figure \ref{pol_mag}, and squares denote the three newly identified condensations. The {\it Right} panel shows the residual derived by subtracting the model from the observation; solid and dotted contours show the positive and negative emissions, respectively, with the contour levels of $\pm0.018, 0.036$~Jy\,beam$^{-1}$. \label{imfit}}
\end{figure}

It has been shown that a star-forming cloud generally collapses to a flattened or filamentary configuration to allow efficient fragmentation to follow \citep{Larson85,Pon11}. The observations of the G35.2N clump is consistent with this theoretical prediction. To further understand the fragmentation property of the clump, it is instructive to compute the thermal Jeans mass, which is about 1.5~$M_{\odot}$ at a density of $1.0\times10^6$~cm$^{-3}$ and a temperature of 30~K. The analyses of the gravitational instability of molecular gas sheets and filaments by Larson (1985) are probably of more relevance to G35.2N. Larson (1985) argued that rotation and magnetic field do not fundamentally change the characteristic mass, $M_{\rm c}(M_{\odot})=2.4T^2/\mu,$ where $T$ is the gas temperature and $\mu$ is the surface density in $M_{\odot}{\rm pc}^{-2}$, of the fragments. For G35.2N, $M_{\rm c}\simeq0.22~M_{\odot}$. Even without taking into account the mass of the forming stars and of the gas already dispersed by the wide-angle outflow, the clump mass is orders of magnitude greater than the Jeans mass or $M_{\rm c}$, and is expected to fragment into hundreds of density enhancements if the dynamics is solely controlled by the gravity and thermal pressure. In Figure \ref{cont}c, there are only six prominent condensations detected at a spatial resolution (0.01~pc) finer than the Jeans length (0.04~pc). We also perform a multi-component Gaussian fitting to the dust emission using the MIRIAD task IMFIT, and identify three new possible condensations, MM1c, MM2c, and MM2d (see Figure \ref{imfit}). Nevertheless, the total number of the condensations is far less than what is expected in thermal fragmentation. Such observational results have been seen in some other high-mass star-forming regions \citep{QZhang09,Qiu11}, and strongly suggest that other mechanisms, e.g., turbulence, magnetic field, and rotation, are playing important roles in the dynamical evolution of the clump. A highly \emph{turbulent} core with a radius of 0.1~pc and a mass of $240~M_{\odot}$, as described by Zhang et al. (2013) in applying the model of McKee \& Tan (2003) to G35.2N, is obviously inconsistent with the high-angular-resolution observations obtained here with the SMA and with the ALMA \citep{Sanchez-Monge13}. Recent numerical simulations of the collapse and fragmentation of magnetized, massive star-forming cores show that cores with $M/{\phi}_B\sim2$ form a single star rather than fragmenting to a small cluster \citep[e.g.,][]{Commercon11,Myers13}. This is inconsistent with our observations either; the measured $M/{\phi}_B$ in G35.2N is $\sim2$--3, and the clump is highly fragmented. We expect that both the B field and turbulence help to increase the effective Jeans mass or the characteristic mass of the fragments, but neither (or their combination) is sufficient to completely suppress fragmentation on $\lesssim0.1$~pc scales.

\subsection{Winding in the deep -- rotationally distorted magnetic field?} \label{winding}
\subsubsection{A schematic picture}
\begin{figure}
\epsscale{1} \plotone{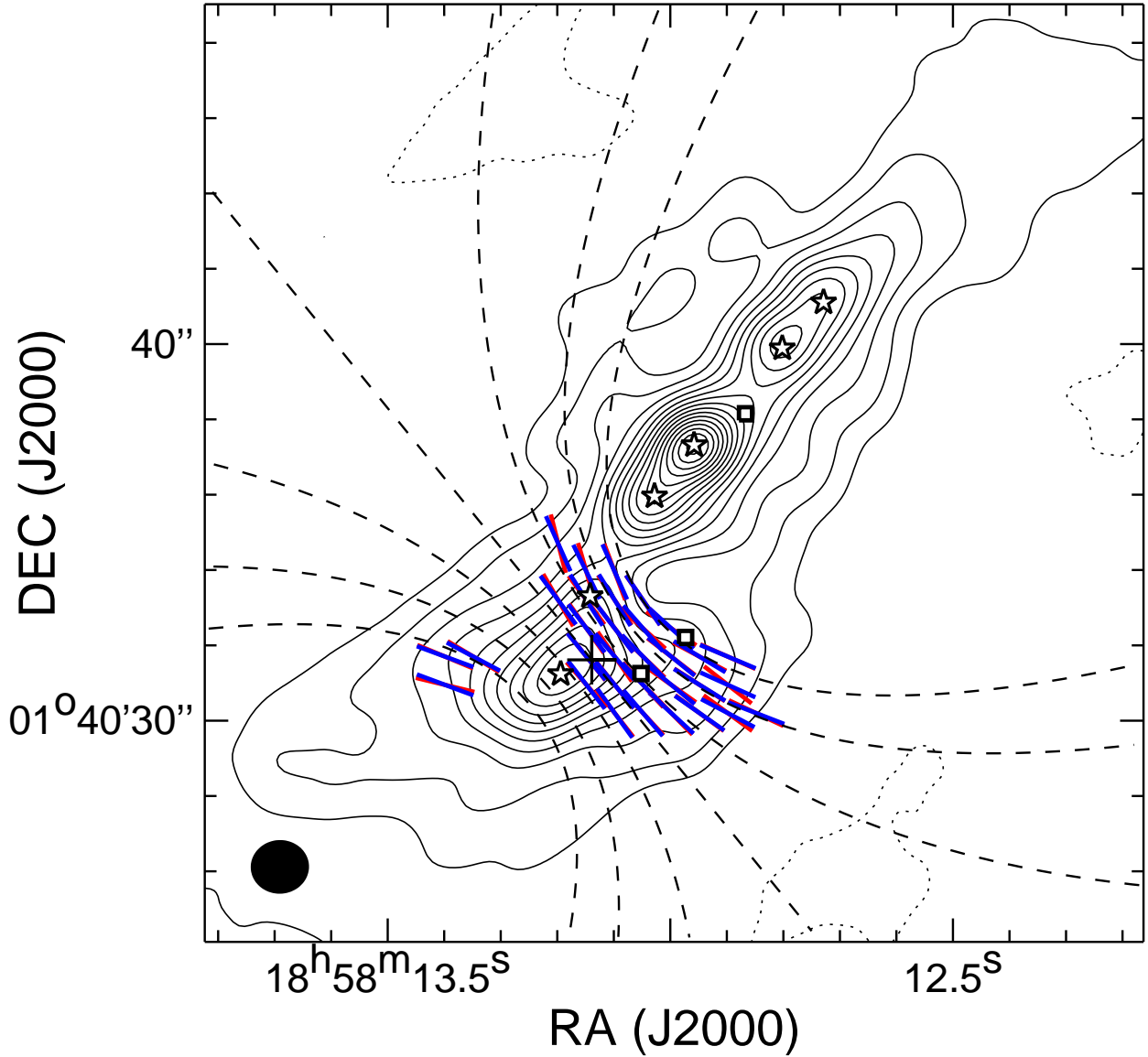}
\caption{Contours are the same as those in Figure \ref{pol_mag}; symbols of stars and squares are the same as those in Figure \ref{imfit}. Red segments show the measured directions of the B field in the southern part of the clump. We fit this part of the B field with a set of parabola, $y-y_0=g_i+g_iC(x-x_0)^2$, following Girart et al. (2006) and Rao et al. (2009), and the best-fit model yields $C=0.053$, the center of symmetry, $(x_0, y_0)$, located at $\rm{(R.A., Decl.)}_{J2000}=(18^h58^m13.\!^s139, +01^d40^m31.\!^s62)$ (denoted as a plus sign), and a position angle of $-51^{\circ}$ for the $y$-axis of the parabola. Dashed curves show a representative set of the fitted parabola. Blue segments depict the tangential directions of the fitted parabola at the positions where the B field directions are measured. \label{hourglass}}
\end{figure}

The sub-parsec clump in G35.2N has been interpreted as a prototype of a massive interstellar disk or toroid, with the ``disk'' being nearly edge-on, having LSR velocities increasing from northwest to southeast \citep{Dent85,Little85,Brebner87,Little98,Lopez-Sepulcre09}. Our observations provide significantly new insights into the dynamics of the clump. Going from the northern to the southern part of the clump, the B field direction takes a $90^{\circ}$ turn and is correlated with the velocity field revealed by the high-density tracing molecular lines (see Section \ref{kinematics}). We also find that the B field in the southern part appears to be slightly pinched and can be fitted with a set of parabola (Figure \ref{hourglass}), which provides marginal evidence for a magnetically regulated collapse of MM2 \citep[e.g.,][]{Girart06,Goncalves08}. In Figure \ref{imfit}, the dust condensations in the northern part are distributed approximately along the major axis of the clump, while those in the southern part have a Trapezium-like distribution. All this suggests that the clump is composed of two dynamically different components rather than coherently rotating as a whole.

One interpretation of the observed velocity gradient in the northern part of the clump is that it arises from the rotation of the dense gas. Considering the orientation of the B field in the cloud revealed by the single-dish observations (see Figure \ref{pol_mag}b), we speculate that the dense clump forms as the cloud collapses more along the direction of the large scale B field. In this dynamical process, the dense gas breaks into two parts. The northern part carries over most of the angular momentum that is not dissipated and forms a fast rotating system. The system could be a rotating toroid embedded with multiple fragments. Or it is a hierarchical structure consisting of two cores (MM1 and MM3) in an orbital motion, and each core further fragments and collapses into a binary or multiple star. In the latter case, the projected separation of $\sim0.05$~pc and the line-of-sight velocity difference of $\sim2.5$~km\,s$^{-1}$ between MM1 and MM3 yield a dynamical mass of $\gtrsim73~M_{\odot}$, which is compatible with the combined mass of the two cores (Table \ref{table2}). The southern part, however, does not participate in the rotating motion and is expected to have a mean velocity similar to that of the parent cloud. This hypothesis is indeed supported by the large scale kinematics seen in $^{13}$CO (1--0) observations obtained from the Galactic Ring Survey, from which an LSR velocity of 35.2~km\,s$^{-1}$ is derived for the ambient cloud of G35.2N \citep[][also see Figure \ref{gradient}]{Jackson06,Roman09}. Also, one may expect that the angular momentum in the northern part has a significant influence on the fragmentation, causing the resulted fragments to be distributed within the rotational plane. The observed morphology of the B field in the clump can be naturally incorporated in the above scenario. Figure 9 provides an overall picture of our interpretations of the B field structure: in the northern part, the B field is pulled by the rotation into a \emph{toroidal} configuration, whereas in the southern part, the B field is only gently squeezed by the collapse and remains to be in a \emph{poloidal} configuration (aligned with the B field in the cloud).

\subsubsection{Quantitative comparisons}
We quantitatively investigate the feasibility of our interpretation of a rotationally distorted B field in the northern part of the clump. According to Matsumoto et al. (2006), a $90^{\circ}$ misalignment between the axis of an outflow and the mean direction of an ordered B field may occur if the rotational energy is at least comparable to the magnetic energy \citep[also see][]{Rao09}. Assuming the measured velocity gradient, ${\Delta}v/L$, in the northern part of G35.2N is arising from a disk-like structure with a diameter $L$, the rotational energy amounts to $1/2(ML^2/8)({\Delta}v/L)^2$, where $M$ is the mass of the rotating structure. The averaged ratio of the rotational energy to the magnetic energy, ${\beta}_{\rm rot}=({\Delta}v)^2/(8V_{\rm A}^2)$, is found to be 0.5--1.3 for an end-to-end velocity difference of $\sim4$~km\,s$^{-1}$ and $B_0\sim1.4$--0.9~mG. Hence the rotational energy does appear to be comparable to the magnetic energy.

Alternatively, if a B field is wound by rotation into an overwhelmingly toroidal configuration, the centrifugal force is presumably overcoming the magnetic force. More quantitatively, Machida et al. (2005) found that whether the centrifugal force or the magnetic force regulates the evolution of a collapsing core can be evaluated by the ratio of the angular velocity to the magnetic field strength, $\omega/B$. If this ratio exceeds a critical value, $0.39\sqrt{G}/c_{\rm s}$, where $c_{\rm s}$ is the sound speed, the centrifugal force dominates the dynamics. For the G35.2N clump, $\omega\sim5.1\times10^{-5}$~year$^{-1}$ and $c_{\rm s}\sim0.33$~km\,s$^{-1}$ at 30~K, leading to $\omega/B\simeq(3.6$--$5.7)\times10^{-8}$ yr$^{-1}$\,$\mu$G$^{-1}$, which is on the same order but lower than the critical value of $9.7\times10^{-8}$ yr$^{-1}$\,$\mu$G$^{-1}$. We stress that both $\omega$ and $c_{\rm s}$ are averaged over the entire clump. Toward the inner part of the clump, $\omega$ increases since the cores appear to rotate faster, and $c_{\rm s}$ also increases since the gas temperature is higher. Thus the measured $\omega/B$ would be greater than the critical value. For example, considering the dense gas in MM1, $\omega$ measures $\sim1.5\times10^{-4}$~year$^{-1}$ (Figure \ref{gradient}f) and $c_{\rm s}$ reaches 0.71~km\,s$^{-1}$; the measured $\omega/B$ is more than 2 times greater than the critical value, confirming that the centrifugal force is dominating over the magnetic force.

The above calculations show that energy-wise, the B field is likely to be significantly distorted by the rotation. But, is there sufficient time for a toroidal B field to develop? The rotation period, $2{\pi}L/{\Delta}v$, is estimated to be $1.2\times10^5$~yr. The dense clump with an averaged density of $1.0\times10^6$~cm$^{-3}$ presumably form from a cloud with a lower density ($\lesssim10^4$~cm$^{-3}$), which has a free-fall time $>10^5$~yr. In addition, with the detection of HMCs and a radio source in MM1, it is feasible to expect that the clump has a dynamical age of order $10^5$~yr \citep[e.g.,][]{Charnley97}. Therefore, it is likely that the rotation has proceeded for a few periods, and significantly distorted the B field into a predominantly toroidal configuration.

Finally, another piece of evidence supporting for the presence of a toroidal B field in the clump comes from polarization observations ($0.\!''2$ resolution) of OH maser emission; Hutawarakorn \& Cohen (1999) detected a few Zeeman pairs within $0.\!''5$ of MM1b and found that the direction of the B field along the line of sight reverses from the southeast to the northwest of MM1b, and the direction reversal was interpreted as being due to a toroidal B field.

\subsubsection{Connecting the B field to the outflow}
\begin{figure}
\epsscale{1} \plotone{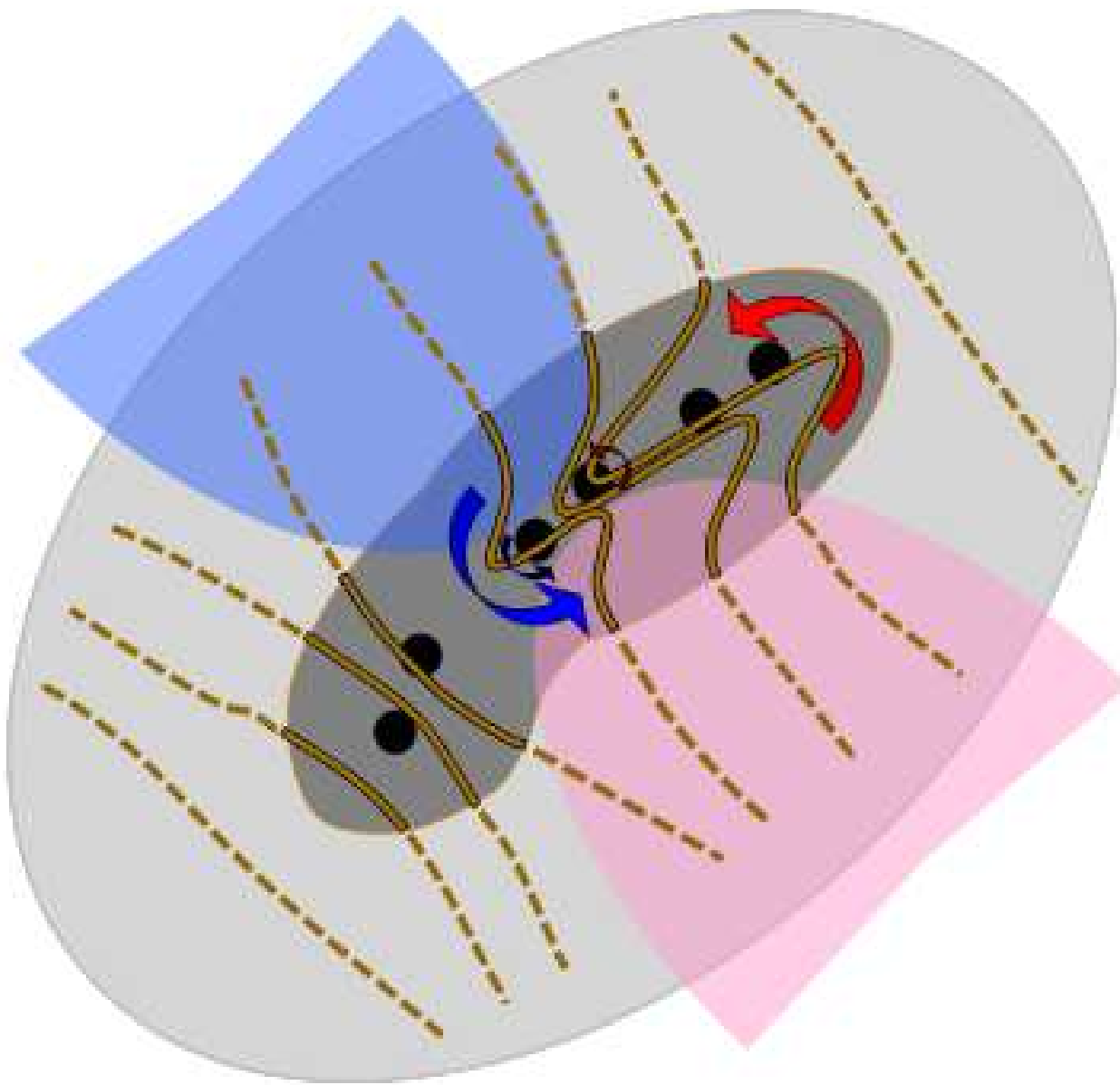}
\caption{A schematic view of our interpretation of the kinematics and magnetic fields in G35.2N. A filamentary structure in dark gray depicts the clump observed with the SMA, and six black dots indicate the deeply embedded dust condensations (see Figure \ref{cont}c). The northern part of the clump is suggested to be rotating, as denoted by curved arrows, whose colors represent the sense of the rotation. A large ellipse filled with light gray shows the parent cloud observable with the JCMT/SCUBA. Two parabolic structures in light blue and red illustrate the blueshifted and redshifted lobes of a bipolar outflow. Solid and dotted lines visualize the magnetic field inside the clump (observed with the SMA) and that outside the clump (inferred from early JCMT observations), respectively. The cartoon is schematic only, and not to be scaled with the observed sizes of the various structures. \label{cartoon}}
\end{figure}

Both well-collimated and wide-angle outflows have been seen in high-mass star-forming regions \citep[e.g.,][]{Qiu08,Qiu09a,Qiu09b,Qiu11a,Qiu12}. Here in G35.2N, the CO (7--6) observations reveal a parsec-sized wide-angle outflow (Figure \ref{outflow}, also see Figure \ref{cartoon}). The outflow appears to originate from one of the two HMCs (MM1a or MM1b, see Section \ref{HMC-outflow}), which must have been undergoing collapse and spun up to form a rotating disk at the center. Indeed, there is evidence for the presence of a rotating disk in MM1b from the ALMA observations \citep{Sanchez-Monge13}, though it is yet to be confirmed whether the outflow is driven from MM1b. In line with our above interpretations, the ambient B field of the inner $\sim0.1$~pc part of the outflow is dominated by a toroidal component, and that component could be further traced down to a 0.01~pc scale taking into account previous OH maser observations \citep{Hutawarakorn99}. Thus the wide-angle outflow is most likely associated with a strong toroidal B field. How is this outflow driven?

Bipolar outflows in protostars or young stars are believed to be magnetically driven. In the most quoted magneto-centrifugal wind theory, the outflowing gas is centrifugally ejected, either from the inner edge of an accretion disk or over a wide range of disk radii, along open poloidal field lines \citep[e.g.,][]{Shu94, Shang06,Pudritz07,Fendt09}. On the other hand, the magnetic tower model suggests that an outflow is accelerated by the magnetic pressure gradient of a wound-up volume of a toroidal B field \citep{Uchida85,Lynden-Bell96,Lynden-Bell03}. Many MHD simulations of collapsing magnetized cores have shown that a bipolar outflow is launched from a strongly rotating region when a predominantly toroidal B field has been built up by the collapse-spin up process \citep[e.g.,][]{Tomisaka98,Matsumoto04,Banerjee06,Banerjee07,Seifried12}. The observations of G35.2N are apparently consistent with such a scenario. Furthermore, numerical simulations often found two-component outflows: a larger, low-velocity, and less collimated outflow emerging earlier than an inner, faster, and jet-like outflow \citep[e.g.,][]{Tomisaka02,Banerjee06,Machida08,Seifried12,Tomida13}. While some claim that the former is a magnetic tower flow and the latter is launched by the magneto-centrifugal force \citep{Banerjee06,Banerjee07}, others argue for an opposite view \citep{Machida08,Tomida13}. The situation can be even more complicated considering possible dependence of the outflow morphology on the B field strength \citep[e.g.,][]{Hennebelle08,Seifried12,Bate13}. In addition, radiation pressure could be important in widening massive outflows \citep{Vaidya11}. Thus we are not able to infer the driving mechanism of the G35.2N outflow solely based on its morphology. Regarding the B field morphology, existing observations suggest a strong toroidal component in the inner part of the outflow, but a poloidal component certainly exists. To determine whether a magnetic tower or the centrifugal force dominates the acceleration of the outflowing gas, one may need a complete knowledge of three-dimensional B field and velocity structures \citep{Seifried12}. Since we are only probing the B field projected in the plane of sky, and the estimate of the B field strength is indirect, it is not possible to compare the toroidal to poloidal components in detail. Hence whether or to what extent a magnetic tower plays a role in accelerating the outflow remains open. But if the growing tower is important, the central collapse-spin up process significantly helps to create at least part of the observed toroidal B field.

\subsection{Infall along a magnetized cylinder?}
Provided the remarkable filamentary morphology of the clump seen in Figure \ref{cont}c, it is likely that we are observing a dense gas cylinder. A gas cylinder may fragment along its major axis due to the ``sausage'' or ``varicose'' instability, and the resulted fragments will have a characteristic spacing \citep[e.g.,][]{Jackson10}. In Fgiure \ref{imfit}a, the dust condensations in the northern part are nearly equally spaced, in qualitative agreement with the fragmentation of a cylinder. Since a cylinder/filament is not expected to be rotating too fast about its minor axis\footnote[9]{Some large and extensively studied filaments are found to exhibit velocity gradients of only $\sim$0.1--1~km\,s$^{-1}$\,pc$^{-1}$ along their major axes.}, the observed velocity gradient ($\sim50$~km\,s$^{-1}$\,pc$^{-1}$) is most likely arising from an infall motion along the cylinder axis rather than from a rotation. From Figure \ref{gradient}d, the infall velocity is $\sim$2/sin$i$~km\,s$^{-1}$, where $i$ is the inclination angle of the cylinder with respective to the plane of sky. For a cylinder with a mass of  $100~M_{\odot}$ (roughly the gas mass of the northern clump plus the mass of the forming stars) and a length of 0.1~pc, the line-of-sight component of the free-fall velocity reaches $\sim4({\rm cos}i)^{1/2}{\rm sin}i$~km\,s$^{-1}$, which is not incompatible (e.g., for $i\simeq54^{\circ}$) with the observed velocity gradient.

However, in the SCUBA map (Figure \ref{cont}a), the clump is much less elongated, and on a larger scale, we do not detect any filamentary structure where the cylinder is embedded. Furthermore, self-gravitating cylinders have a critical linear mass density, $2v^2/G$, where $v$ is the sound speed $c_{\rm s}$ in case of thermal support \citep{Stodolkiewicz63,Ostriker64} or the turbulent velocity dispersion ${\delta}V_{\rm los}$ in case of turbulent support \citep{Fiege00}. Above the critical value, the cylinder would radially collapse into a line. The B field can somewhat increase the critical linear mass density, but probably by small factors $\sim1$ \citep{Fiege00}. For G35.2N, the measured linear mass density is of order $1000~M_{\odot}{\rm pc}^{-1}$ (without taking into account the mass of the forming stars and of the gas already dispersed by the outflow), which is about two times greater than the critical value of $470~M_{\odot}{\rm pc}^{-1}$. Unless the cylinder is largely inclined to the line of sight (i.e., $i>60^{\circ}$), the linear mass density seems to be too high to allow a stable cylinder to exist. Also, the southern part of the clump, which appears very different from the northern part in the B field, kinematics, and fragmentation properties, cannot be easily incorporated in a unified picture of a cylinder.  Considering all these difficulties in understanding the observed clump as a cylinder, the interpretation of a rotating system (Section \ref{winding}) for the northern part of the clump appears more robust than that of a collapsing cylinder.

\section{Conclusions}
With the sensitive and high-angular-resolution observations made with the SMA, we detect a well-ordered B field threading G35.2N, which is an apparently filamentary, massive, and cluster-forming clump. Based on a statistical analysis of the observed polarization dispersion, we derive a B field strength of $\sim1$~mG, a turbulent-to-magnetic energy ratio of order unity, and a mass-to-magnetic flux ratio of $\sim2$--3 times the critical value.

The B field morphology is found to vary depending on the kinematics revealed by high-density tracing molecular lines. In the northern part of the clump, which exhibits a velocity gradient, the B field is largely aligned with the long axis of the clump (hence follows the velocity gradient). Approaching the southern part, which has a relatively uniform velocity field, the B field takes a great turn of $90^{\circ}$ to a direction perpendicular to the clump elongation. Although we cannot rule out the possibility that the velocity gradient is due to an infall motion along a gas cylinder, a synthesis of all the available observations prefers a rotating motion to be the origin of the velocity gradient. Consequently the B field in the northern part is pulled by the centrifugal force into an overwhelmingly toroidal configuration, whereas the B field in the southern part remains in a poloidal configuration, that is, aligned with a large scale B field which guides the collapse of the cloud from which the clump forms. Therefore we are likely witnessing a transition from a poloidal to toroidal configuration for the B field in this region. Detailed calculations adopting the B field strength derived from the statistical analysis supports such a scenario. Our observations also provide evidence for a wide-angle outflow driven from a strongly rotating region whose B field is largely toroidal.

Last, but not least, the supercritical clump is observed to be significantly fragmenting, but to a degree far from expected for a purely thermal fragmentation. Our observations and calculations suggest that both the B field and turbulence, as well as the potential rotation, all play a role in interplaying with the gravity and shaping the dynamical evolution of the clump.

\acknowledgments Part of the data were obtained in the context of the SMA legacy project: ``Filaments, Magnetic Fields, and Massive Star Formation'' (PI: Qizhou Zhang). We acknowledge all the members of the SMA staff who made these observations possible. The JCMT SCUBA data were obtained from the JCMT archive (Program ID: M97BH07). This research used the facilities of the Canadian Astronomy Data Centre operated by the National Research Council of Canada with the support of the Canadian Space Agency. Part of this research was undertaken when KQ was a postdoctoral fellow at Max-Planck-Institut fuer Radioastronomie. KQ is supported by the 985 project of Nanjing University. QZ is partially supported by the NSFC grant 11328301.

\clearpage

\begin{deluxetable}{llcllll}
\tablecaption{List of Observational Parameters \label{table1}}
\tablehead{
\colhead{Date of} & \colhead{Configuration} & \colhead{Number of} &
\colhead{${\tau}_{\rm 225}$} & \colhead{Bandpass\tablenotemark{a}} & \colhead{Gain} &
\colhead{Flux} \\
\colhead{Observations} & \colhead{} & \colhead{Antennas} & \colhead{} &
\colhead{Calibrator} & \colhead{Calibrator} & \colhead{Calibrator}
}
\startdata
2010 Oct. 05 & Compact    & 7 & 0.05 & 3C273  & J1751+096 & Titan    \\
             &            &   &      &        & J1830+063 &          \\
2011 Jul. 02 & Subcompact & 7 & 0.07 & 3C279  & J1751+096 & Callisto \\
             &            &   &      &        & J1733-130 &          \\
2011 Jul. 21 & Extended   & 7 & 0.05 & 3C279  & J1751+096 & Callisto \\
             &            &   &      &        & J1733-130 &          \\
2011 Jul. 23 & Extended   & 8 & 0.06 & 3C279  & J1751+096 & Callisto \\
2012 Mar. 28 & Extended   & 6 & 0.07 & 3C279  & J1751+096 & Titan    \\
\enddata
\tablenotetext{a}{Also used for instrumental polarization (leakage) calibrations.}
\end{deluxetable}

\begin{deluxetable}{cccllrll}
\tablewidth{0pc} \tablecaption{Measured and computed parameters of the three dust cores \label{table2}}
\tablehead{ \colhead{} & \colhead{R.A.} & \colhead{Decl.} & \colhead{Peak Flux} &
\colhead{Total Flux} & \colhead{Temp.} & \colhead{$\beta$\tablenotemark{c}} & \colhead{Mass} \\
\colhead{} & \colhead{(J2000)} & \colhead{(J2000)} & \colhead{(Jy\,beam$^{-1}$)} & \colhead{(Jy)} & \colhead{(K)} & \colhead{} &
\colhead{($M_{\odot}$)} }

\startdata  MM1 & $18^{\mathrm h}58^{\mathrm m}12.98^{\mathrm s}$ & $1^{\circ}40'36.8''$ & 1.2  & 3.0 & 50--135\tablenotemark{a} & 1.3 & 25--8.1 \\
                  MM2 & $18^{\mathrm h}58^{\mathrm m}13.18^{\mathrm s}$ & $1^{\circ}40'31.6''$ & 0.95 & 3.5 & 30\tablenotemark{b}  & 1.5 & 71 \\
                  MM3 & $18^{\mathrm h}58^{\mathrm m}12.77^{\mathrm s}$ & $1^{\circ}40'40.2''$ & 0.69 & 1.7 & 30\tablenotemark{b} & 2.0 & 65 \\
\enddata
\tablenotetext{a}{The upper limit is from our LTE fitting to the CH$_3$CN (19--18) emission, and the lower limit is from recent NH$_3$ observations \citep{Codella10}.}
\tablenotetext{b}{From early NH$_3$ (1,1), (2,2) observations \citep{Little85}.}
\tablenotetext{c}{Derived by comparing the 340 GHz and 225 GHz continuum maps constructed with the same $(u,v)$ range.}
\end{deluxetable}

\begin{deluxetable}{cccccc}
\tablewidth{0pc} \tablecaption{Results from the statistical analysis of the P.A. dispersion\tablenotemark{a}}
\tablehead{ \colhead{$\delta$} & \colhead{$N$} & \colhead{${\langle}B_t^2\rangle/{\langle}B_0^2\rangle$} & \colhead{${\beta}_{\rm turb}$} & \colhead{$B_0$} & \colhead{$M/{\phi}_B$} \\
\colhead{(mpc)} & \colhead{} & \colhead{} & \colhead{} & \colhead{(mG)} & \colhead{($1/2\pi\sqrt{G}$)} }

\startdata  15.4 & 3.0--8.4 & 0.23--0.64 & 0.7--1.9 & 1.4--0.9 & 1.9--3.2 \\
\enddata
\tablenotetext{a}{Except $\delta$, all the derived parameters depend on the effective depth, $\Delta'\sim8$--$22''$.}
\end{deluxetable}


\begin{thebibliography}{}
\bibitem[Banerjee \& Pudritz 2006]{Banerjee06}Banerjee, R. \& Pudritz, R. 2006, \apj, 641, 949
\bibitem[Banerjee \& Pudritz 2007]{Banerjee07}Banerjee, R. \& Pudritz, R. 2007, \apj, 660, 479
\bibitem[Basu \& Mouschovias 1994]{Basu94}Basu, S. \& Mouschovias, T. Ch. 1994, \apj, 432, 720
\bibitem[Bate et al. 2013]{Bate13}Bate, M. R., Tricco, T. S., \& Price, D. J. 2013, \mnras, in press, arXiv: 1310.1092
\bibitem[Birks et al. 2006]{Birks06}Birks, J. R., Fuller, G. A., \& Gibb, A. G. 2006, \aap, 458, 181
\bibitem[Brebner et al. 1987]{Brebner87}Brebner, G. C., Heaton, B., Cohen, R. J., \& Davies, S. R. 1987, \mnras, 229, 679
\bibitem[Brown et al. 1982]{Brown82}Brown, A. T., Little, L. T., MacDonald, G. H., Matheson, D. N. 1982, \mnras, 201, 121
\bibitem[Chandrasekhar \& Fermi 1953]{Chandrasekhar53}Chandrasekhar, S. \& Fermi, E. 1953, \apj, 118, 113
\bibitem[Charnley 1997]{Charnley97}Charnley, S. B. 1997, \apj, 481, 396
\bibitem[Codella et al. 2010]{Codella10}Codella, C., Cesaroni, R., L\'{o}pez-Sepulcre, A., et al. 2010, \aap, 510, A86
\bibitem[Commer\c{c}on et al. 2011]{Commercon11}Commer\c{c}on, B., Hennebelle, P., \& Henning, T. 2011, \apj, 742, 9
\bibitem[Crutcher 2012]{Crutcher12}Crutcher, R. M. 2012, \araa, 50, 29
\bibitem[Dent et al. 1985]{Dent85}Dent, W. R. F., Little, L. T., Kaifu, N., Ohishi, M., Suzuki, S. 1985, \mnras, 146, 375
\bibitem[Dent et al. 1989]{Dent89}Dent, W. R. F., Sandell, G., Duncan, W. D., \& Robson, E. I. 1989, \mnras, 238, 1497
\bibitem[Fendt 2009]{Fendt09}Fendt, C. 2009, \apj, 692, 346
\bibitem[Elmegreen 2000]{Elmegreen00}Elmegreen, B. G. 2000, \apj, 530, 277
\bibitem[Fiege \& Pudritz 2000]{Fiege00}Fiege, J. D. \& Pudritz, R. E. 2000, \mnras, 311, 85
\bibitem[Gibb et al. 2003]{Gibb03}Gibb, A. G., Hoare, M. G., Little, L. T., \& Wright, M. C. H. 2003, \mnras, 339, 1011
\bibitem[Girart et al. 2009]{Girart09}Girart, J. M., Beltr\'{a}n, M. T., Zhang, Q., Rao, R., \& Estalella, R. 2009, Science, 324, 1408
\bibitem[Girart et al. 2013]{Girart13}Girart, J. M., Frau, P., Zhang, Q., et al. 2013, \apj, 772, 69
\bibitem[Girart et al. 2006]{Girart06}Girart, J. M., Rao, R., \& Marrone, D. P. 2006, Science, 313, 812
\bibitem[Gon\c{c}alves et al. 2008]{Goncalves08}Gon\c{c}alves, J., Galli, D., Girart, J. M. 2008, \aap, 490, L39
\bibitem[Hartmann et al. 2012]{Hartmann12}Hartmann, L., Ballesteros-Paredes, J., \& Heitsch, F. 2012, \mnras, 420, 1457
\bibitem[Hennebelle et al. 2011]{Hennebelle11}Hennebelle, P., Commer\c{c}on, B., Joos, M., et al. \& Teyssier, R. 2011, \aap, 528, 72
\bibitem[Hennebelle \& Fromang 2008]{Hennebelle08}Hennebelle, P. \& Fromang, S. 2008, \aap, 477, 9
\bibitem[Hildebrand 1983]{Hildebrand83}Hildebrand, R. H. 1983, \qjras, 24, 267
\bibitem[Hildebrand et al. 2009]{Hildebrand09}Hildebrand, R. H., Kirby, L., Dotson, J. L., Houde, M., \& Vaillancourt, J. E. 2009, \apj, 696, 567
\bibitem[Houde et al. 2009]{Houde09}Houde, M., Vaillancourt, J. E., Hildebrand, R. H., Chitsazzadeh, S., \& Kirby, L. 2009, \apj, 706, 1504
\bibitem[Hutawarakorn \& Cohen 1999]{Hutawarakorn99}Hutawarakorn, B. \& Cohen, R. J. 1999, \mnras, 303, 845
\bibitem[Jackson et al. 2006]{Jackson06} Jackson, J.~M., et al. 2006, \apjs, 163, 145
\bibitem[Jackson et al. 2010]{Jackson10}Jackson, J. M., Finn, S. C., Chambers, E. T., Rathborne, J. M., \& Simon, R. 2010, \apj, 719, L185
\bibitem[Koch et al. 2010]{Koch10}Koch, P. M., Tang, Y.-W., \& Ho, P. T. P. 2010, \apj, 721, 815
\bibitem[Larson 1985]{Larson85}Larson, R. B. 1985, \mnras, 214, 379
\bibitem[Li et al. 2010]{Li10}Li, H.-B., Houde, M., Lai, S.-P., \& Sridharan, T. K. 2010, \apj, 718, 905
\bibitem[Lynden-Bell 1996]{Lynden-Bell96}Lynden-Bell, D. 1996, \mnras, 279, 389
\bibitem[Lynden-Bell 2003]{Lynden-Bell03}Lynden-Bell, D. 2003, \mnras, 341, 1360
\bibitem[Little et al. 1985]{Little85}Little, L. T., Dent, W. R. F., Heaton, B., Davies, S. R., \& White, G. J. 1985, \mnras, 217, 227
\bibitem[Little et al. 1998]{Little98}Little, L. T., Kelly, M. L., \& Murphy, B. T. 1998, \mnras, 294, 105
\bibitem[Liu et al. 2013]{Liu13}Liu, H. B., Qiu, K., Zhang, Q., Girart, J. M., \& Ho, P. T. P. 2013, arXiv:1305.3681
\bibitem[L\'{o}pez-Sepulcre et al. 2009]{Lopez-Sepulcre09}L\'{o}pez-Sepulcre, A., Codella, C., Cesaroni, R., Marcelino, N., \& Walmsley, C. M. 2009, \aap, 499, 811
\bibitem[Mac Low \& Klessen 2004]{MacLow04}Mac Low, M.-M. \& Klessen, R. S. 2004, Review of Modern Physics, 76, 125
\bibitem[Machida et al. 2005]{Machida05}Machida, M. N., Matsumoto, T., Tomisaka, K., \& Hanawa, T. 2005, \mnras, 362, 369
\bibitem[Machida et al. 2008]{Machida08}Machida, M. N., Inutsuka, S. I., \& Matsumoto, T. 2008, \apj, 676, 1088
\bibitem[Marrone \& Rao 2008]{Marrone08}Marrone, D. P. \& Rao, R. 2008, Proc. SPIE, 7020
\bibitem[Matsumoto et al. 2006]{Matsumoto06}Matsumoto, T., Nakazato, T., Tomisaka, K. 2006, \apj, 637, L105
\bibitem[Matsumoto \& Tomisaka 2004]{Matsumoto04}Matsumoto, T. \& Tomisaka, K. 2004, \apj, 616, 266
\bibitem[McKee \& Tan 2003]{McKee03}McKee, C. F. \& Tan, J. C. 2003, \apj, 585, 850
\bibitem[Mouschovias et al. 2006]{Mouschovias06}Mouschovias, T. Ch., Tassis, K., \& Kunz, M. W. 2006, \apj, 646, 1043
\bibitem[Myers et al. 2013]{Myers13}Myers, A. T., McKee, C. F., Cunningham, A. J., Klein, R. I., \& Krumholz, M. R. 2013, \apj, 766, 97
\bibitem[Nakano \& Nakamura 1978]{Nakano78}Nakano, T. \& Nakamura, T. 1978, \pasj, 30, 681
\bibitem[Ostriker et al. 2001]{Ostriker01}Ostriker, E. C., Stone, J. M., \& Gammie, C. F. 2001, \apj, 546, 980
\bibitem[Ostriker 1964]{Ostriker64}Ostriker, J. 1964, \apj, 140, 1056
\bibitem[Padoan et al. 2001]{Padoan01}Padoan, P., Goodman, A., Draine, B. T., et al. 2001, \apj, 559, 1005
\bibitem[Peters et al. 2011]{Peters11}Peters, T. Banerjee, R., Klessen, R. S., \& Mac Low, M.-M. 2011, \apj, 729, 72
\bibitem[Pon et al. 2011]{Pon11}Pon, A., Johnstone, D, \& Heitsch, F. 2011, \apj, 740, 88
\bibitem[Pudritz et al. 2007]{Pudritz07}Pudritz, R., Ouyed, R., Fendt, Ch., \& Brandenburg, A. 2007, in Protostars and Planets V, ed. B. Reipurth et al. (Tucson: Univ. Arizona Press)
\bibitem[Qiu et al. 2008]{Qiu08}Qiu, K., et al. 2008, \apj, 685, 1005
\bibitem[Qiu et al. 2011a]{Qiu11a}Qiu, K., Wyrowski, F., Menten, K. M., et al. 2011a, \apj, 743, L25
\bibitem[Qiu \& Zhang 2009]{Qiu09a}Qiu, K. \& Zhang, Q. 2009, \apj, 702, L66
\bibitem[Qiu et al. 2012]{Qiu12}Qiu, K., Zhang, Q., Beuther, H., \& Fallscheer, C. 2012, 756, 170
\bibitem[Qiu et al. 2009]{Qiu09b}Qiu, K., Zhang, Q., Wu, J., \& Chen, H.-R. 2009, \apj, 696, 66
\bibitem[Qiu et al. 2011b]{Qiu11b}Qiu, K., Zhang, Q., \& Menten, K. M. 2011b, \apj, 728, 6
\bibitem[Rao et al. 2009]{Rao09}Rao, R., Girart, J. M., Marrone, D. P., Lai, S.-P., \& Schnee, S. 2009, \apj, 707, 921
\bibitem[Roman-Duval et al. 2009]{Roman09}Roman-Duval, J., Jackson, J. M., Heyer, M., et al. 2009, \apj, 699, 1153
\bibitem[S\'{a}nchez-Monge et al. 2013]{Sanchez-Monge13}S\'{a}nchez-Monge, \'{A} et al. 2013, \aap, 552, L10
\bibitem[Seifried et al. 2012]{Seifried12}Seifried, D., Pudritz, R. E., Banerjee, R., Duffin, D., \& Klessen, R. S. 2012, \mnras, 422, 347
\bibitem[Shang et al. 2006]{Shang06}Shang, H., Allen, A., Li, Z.-Y., et al. 2006, \apj, 649, 845
\bibitem[Shu et al. 1987]{Shu87}Shu, F. H., Adams, F. C., \& Lizano, S. 1987, \araa, 25, 23
\bibitem[Shu et al. 1994]{Shu94}Shu, F., Najita, J., Ostriker, E., et al. 1994, \apj, 429, 781
\bibitem[Stod\'{o}lkiewicz 1963]{Stodolkiewicz63}Stod\'{o}lkiewicz, J. S. 1963, Acta Astron., 13, 30
\bibitem[Tang et al. 2013]{Tang13}Tang, Y.-W., Ho, P. T. P., Koch, P. M., Guilloteau, S., Dutrey, A. 2013, \apj, 763, 135
\bibitem[Tassis et al. 2009]{Tassis09}Tassis, K., Dowell, C. D., Hiledebrand, R. H., Kirby, L., \& Vaillancourt, J. E. 2009, \mnras, 399, 1681
\bibitem[Tomida et al. 2013]{Tomida13}Tomida, K. et al. 2013, \apj, 763, 6
\bibitem[Tomisaka 1998]{Tomisaka98}Tomisaka, K. 1998, \apj, 502, 163
\bibitem[Tomisaka 2002]{Tomisaka02}Tomisaka, K. 2002, \apj, 575, 306
\bibitem[Uchida \& Shibata 1985]{Uchida85}Uchida, Y. \& Shibata, K. 1985, \pasj, 37, 515
\bibitem[Vaidya et al. 2011]{Vaidya11}Vaidya, B., Fendt, C., Beuther, H., \& Porth, O. 2011, \apj, 742, 56
\bibitem[Vall\'{e}e \& Bastien 2000]{Vallee00}Vall\'{e}e, J. P. \& Bastien, P. 2000, \apj, 530, 806
\bibitem[Zhang et al. 2009a]{Zhang09}Zhang, B., Zheng, X.-W., Reid, M. J., et al. 2009a, \apj, 693, 419
\bibitem[Zhang et al. 2009b]{QZhang09}Zhang, Q., Wang, Y., Pillai, T., \& Rathborne, J. 2009b, \apj, 696, 268
\bibitem[Zhang et al. 2013]{Zhang13}Zhang, Y. et al. 2013, \apj, 767, 58
\end{thebibliography}
\end{document}